**The complex world of honey bee vibrational signaling: A response to Ramsey et al. (2017)**

**Heather C. Bell[1]\*, Parry M. Kietzman[2], and James C. Nieh[1]\***

\*Corresponding authors
hebell@ucsd.edu
jnieh@ucsd.edu

[1]UCSD
Division of Biological Sciences
Section of Ecology, Behavior, and Evolution
9500 Gilman Drive, MC0116
La Jolla, CA, 92093
USA

[2]Appalachian Headwaters
PO Box 1500
Lewisberg, WV, 24901
USA

Ramsey et al. [1] report on the characteristics and temporal distribution of an interesting vibrational signal that they term the "whooping signal", primarily based upon a long-term study of vibrations recorded by accelerometers placed inside two honey bee (*Apis mellifera*) colonies, one in France and one in the United Kingdom. The goal of the study, the long-term automated recording and analysis of honey bee vibrational signaling, is worthwhile—but we believe that some of the conclusions drawn by the authors are not well supported, given the evidence.

Honey bees, particularly the relatively well-studied species, *A. mellifera*, have a wide variety of vibrational signals that play an important role in organizing and coordinating colony life [2–13]. Understanding the differences between these signals is important. The crux of our criticism rests on how the authors were able to accurately identify and describe signals based largely upon accelerometers implanted inside colonies without additional behavioral data and using signal processing techniques that may not have been sufficient to discriminate between multiple types of vibrational signals. We also believe that they have incorrectly dismissed an interesting and plausible explanation for their results.

The authors describe the discovery of a new signal, which they call the whooping signal. They state that the whooping signal is "a honeybee vibrational pulse with the same characteristics of what has previously been described as a stop signal" (p 1, [1]). To support the need for this new signal name, they cite different descriptions of what is generally now called the "stop signal" [12,14–16]. The stop signal is a brief vibrational signal, often delivered by waggle dance followers head butting waggle dancers. The stop signal was originally called a begging signal [17] but only rarely elicits food samples [14,18]. However, it is agreed to cause momentary freezing of the signal recipient [3], hence the name "stop signal". Stop signals have also been called a "brief piping signal" or a "nectar forager pipe" [18,19]. Multiple studies now show that stop signals decrease the number of dance circuits performed by a waggle dancing recipient, thereby inhibiting recruitment [14–16,18,20], although other functions are certainly possible. Ramsey et al. [1] note that the stop signal has been implicated in re-allocation of tasks within the hive [20] because it is produced by tremble dancers in response to long nectar unloading wait times. With the exception of its possible role in task allocation, all of the other published and currently accepted functions of the stop signal in *A. mellifera* are consistent with waggle dance inhibition.

Thus, the objections raised by Ramsey et al. [1] about multiple stop signal contexts and functions (danger at a resource during foraging [15], crowding at a resource during foraging [21], and during consensus-forming while house-hunting [16]) are not problematic. In fact, the stop signal can be seen as an elegant evolutionary solution that counterbalances the positive feedback of the waggle dance, which itself serves multiple functions by recruiting nestmates for diverse resources: water, nectar, pollen, resin, nest sites, and, in *Apis dorsata*, migration direction [22–26]. We agree that there is much to be learned about honey bee vibrational signaling and also about the stop signal. However, it is incorrect to imply that the descriptions of what has been called the begging signal or stop signal describe vastly differing phenomena lacking any unifying theme. Below, we detail our main concerns about Ramsey et al. [1].

### *Automated detection*

To identify whooping signals in their hives, the authors created an automated detection algorithm based on a discriminant function that they trained to distinguish whooping signals from non-whooping signals. The authors state that they trained their discriminant function in a semi-supervised fashion, employing discriminant listening to ensure that only "real" whooping signals were classified as whooping signals. However, nowhere in the main paper or the supplementary materials do the authors indicate how they initially parameterized their discriminant function, nor is it indicated how they learned to distinguish whooping from non-whooping signals.

It is not clear, for example, if the authors incorporated the considerable natural variation in the fundamental frequency and durations of stop signals [19,21] into their detection model. Further, they make a problematic assumption that all stop signals have the same number of detectable harmonics, stating that a "common feature of this signal is to have two well pronounced upper harmonics at twice and three times the value of the fundamental frequency. Numerous harmonics at even higher frequencies can also be seen, with negligible but measurable amplitudes" (p 6 [1]). In fact, there is variation in the number of reported harmonics[3,19]. Thom et al. [19] used a microphone to measure the airborne sounds generated by stop signals, and recorded far more non-negligible harmonics (see their Fig. 2). More concerning is the violation of a well-known physical principle: substrate transmission and the exact placement and type of detection devices can dramatically change the amplitudes of measured harmonics [27]. Ramsey et al. [1] parameterized their detections (obtained from recording with accelerometers embedded inside the combs of full colonies) with data from Michelsen et al. [3]. However, Michelsen et al. [3] made their measurements using laser Doppler vibrometry with a small observation hive.

We are also concerned about the issue of vibrational noise, which can be substantial inside honey bee colonies, and are not certain that this technique is effective for recording stop signals. Michelsen reported that the stop signal is attenuated rapidly in the comb [3] and therefore used laser Doppler vibometry to record signals close to the signaler. Using small, three-frame observation hives, we have attempted to record these vibrational signals with sensitive accelerometers embedded into the comb while also scanning the comb with a pressure-condenser microphone. Although pressure microphones scanning over the comb surface clearly detected stop signals, simultaneous accelerometer recordings detected no stop signals and spectral analyses of the accelerometer recordings yielded results similar to recording white noise [28].

Ramsey et al. [1] report that the whooping signal has a fundamental frequency of 355 Hz, and a mean duration of 60 ms (p 6, [1]). However, in the text of paper, Michelsen et al. [3] reported a fundamental frequency of 320 Hz and a duration of "up to" 100 ms. These values agree with the spectrum and waveform of Michelsen's example begging signal [3]. However, a search of the literature yields other descriptions of the stop signal fundamental frequencies. Schlegel et al. [10] reported that stop signals have a fundamental frequency of 407 Hz and a pulse duration of 174±64 ms (mean±1 standard error) for stop signals. Thom et al. [19] recording brief piping calls produced by bees foraging at a nectar feeder when they began to experience deteriorating foraging conditions, as has been described for stop signals [14,18]. Thus, these signals are likely stop signals and share characteristics with stop signals described by other authors [14,18]. Thom et al. [19] reported fundamental frequencies ranging from 270-540 Hz and a duration of 230±100 ms (mean±1 standard deviation). ,14,18]), and Thom et al.[19] reported fundamental frequencies of 270, 380, and 540 Hz at different times within the amplitude modulated stop signal. These results highlight the variation that exists and should be accounted for in the authors' discriminant function.

The authors state that they have distinguished between different types of worker piping, queen piping, rain drops, and whooping signals, but provide no evidence of how this was done. An examination of the data that they used to draw their conclusions shows that vibrations generated by rain drops striking the colony were explicitly excluded. They do not state the specific parameters used to characterize different types of worker piping and queen piping. In addition to these signals, other signals should be discriminated. Multiple kinds of vibrational piping-type signals have been described in honey bees in general [29–32], and in *A. mellifera* specifically [2–13,19,33]. The authors give no indication of how these signals or other vibrations (such as those produced during waggle dancing [34]) were handled.

The authors write that "limited tests strongly support the fact that the vast majority of pulses that we have analysed are exclusively whooping signals as understood by experts," (p 6, [1]). However, no prior studies have identified a honey bee whooping signal. If the authors are stating that the majority of their detections are of stop signals, we respectfully disagree. We find, based upon our experience of hundreds of hours of field observations and listening to thousands of stop signals, that the majority of signals they have identified (their audio of S6 Movie [1]) do not sound like stop signals at all. However, we agree that there is a need to discriminate rigorously between these different signals. For a start, authors, including ourselves, should supply recordings of these sounds and vibrations. We hope to begin remedying this deficit by supplying stop signal wav files in future manuscripts. A good video example with sound is available (https://www.youtube.com/watch?v=VzPpc2zNsz0), and we plan to add more. Detailed, comparative analyses of these and other honey bee vibrational signals would be beneficial.

***Signals have both acoustic and behavioral characteristics***

Aside from their acoustic characteristics, honey bee vibrational signals are also defined behaviorally, and by the context of sender and receiver [14,21]. Focusing solely on an acoustic description, although an important step in signal identification, is not sufficient. Even if signals with the correct acoustic properties were detected, the only way to verify that they are stop signals is to observe the behavior and confirm head-butting—either of another bee or of the substrate. Head-butting is a key component of all published descriptions of the stop signal [10,14,16,18,21,35,36]. Ramsey et al. [1] note that their supplementary videos mostly depict a puzzling lack of any identifiable behavior correlated with the whooping signal (p 16, [1]).

One consequence of having no behavioral context is the collision hypothesis, in which the authors, based upon little evidence (two videos: S4 Movie and S5 Movie), suggest that stop signals convey no information, and are the result of accidental "collisions" between bees (p 16, [1]).

In general, we are concerned about other statements made without or with similarly very weak evidence. The authors state that they detect the whooping signal when "there is no waggle dance or trophallaxis present on the [video] frame under investigation" (p 16, [1]). This is an important assertion, and needs to be documented. Likewise, the authors report that knocking gently on the hive also results in whooping signals. Honey bees will produce piping-like vibrations when struck, pressed down, or otherwise disturbed [37] —reportedly even arising from the heat or electromagnetic radiation generated by a cell phone apparatus [38]. However, to state that these phenomena, produced when a bee is trying to escape or is disturbed, is a signal, and to suggest that similar vibrational sound pulses have no meaning (p 16, [1]), is an overreach, particularly when the authors have not determined which bees are producing the vibrations recorded in their study or the results of these vibrations.

Demonstrating that whooping signals are produced in response to stressful events as described above would require a simple experiment, but instead of doing this (e.g., counting signals before, during, and after knocking the hive), the authors elected to only conduct long-term recordings. The omission of such a simple test leads us to wonder why, then, the authors mention it at all, and also if the signal that they recorded is actually distinct from the stop signal that others have studied, in which the sender generally directs the signal at a specific waggle dancer [15,16].

The authors also state that trophallaxis was sometimes but not always observed related to the whooping signals they were hearing. Although it was originally thought (e.g., [17,22]) that stop signals elicited trophallaxis, this rarely occurs: 15% of cases in [14] and 0% in [18]. If trophallaxis is pertinent to the whooping signal, then it should be quantified.

We are concerned that these videos provide only anecdotal evidence and provide, in several cases, limited video quality that does not illuminate the claims made. They are not a basis for drawing strong conclusions. A more rigorous behavioral analysis would have been beneficial.

Based upon these observations, the authors claim that that results "suggest that this vibrational pulse is generated under many different circumstances, thereby unifying previous publications' conflicting definitions" (p 1, [1]). However, as was previously mentioned, studies on the stop signal are less conflicting than Ramsey et al. [1] imply.

### *Are whooping sounds warm-up vibrations or warm-up signals?*

We suggest an alternative explanation for their results that is also quite interesting. The authors may have recorded warming buzzes, which are produced by the same thoracic muscles as stop signals [3] and therefore share similar characteristics. The authors state that the majority of whooping signals occur at night, with a 50% reduction at around midday (p 13, [1]), that signals are very frequent during the winter months (p 14, [1]), that there was a lack of daytime signaling, except on days of heavy rain (p 14, [1]), that signals increase after the last swarm and in the winter, and that roughly twice as many signals were detected at the periphery of the frame as compared to the center in both the French and UK data sets (p 14, [1]). We note that temperatures typically decrease at night, are lower during the winter, are lower when it is raining, and should be cooler at the periphery than at the center of the comb. In fact, the authors report that "there is a negative effect of outside temperature ($B$ = -0.16, $p$ <0001) on whooping signal occurrences with the greatest number of whooping signals being recorded at temperatures around 0" (p 11, [1]).

Honey bees are known to produce such warm-up vibrations. Seeley and Tautz [5] present an excellent study on how swarms warm up before they take off for flight, producing worker piping that rises up to 200-250 Hz and allows bees to reach the critical temperature for flight. The signals recorded by Ramsey et al. [1] have a different fundamental frequency and may therefore be different than the piping recording by Seeley and Tautz [5], but this is not surprising given that Ramsey et al. [1] were not studying swarms preparing for lift-off. However, some form of vibration generated as a side effect of warming up or a signal that is part of colony thermoregulation appears to account for much of their results.

The authors do consider the warming hypothesis, indicating that their temperature data support it (p 14 [1]). However, they go on to claim that (1) on the coldest day they recorded (December 3), there was a decrease in whooping signals, (2) an increase in signaling in May corresponded to an exceptionally mild spring (daily temperatures ranging from 15 to 28 ℃), and (3) the coldest daily temperatures are known to usually take place at sunrise, and the maxima for average whooping signals over the course of a day (their Fig. 4) occurred closer to the middle of the night (p 14, [1]). With respect to the first point, one outlier in a data set of 261 days of recordings does not provide strong evidence against the overall negative relationship between whooping signal occurrence and temperature. Biological data are inherently noisy, and the utility of collecting large amounts of data is precisely so that any one outlier cannot artificially mask an underlying pattern

[39,40]. Similarly, the second point also does not constitute a genuine refutation of the relationship – particularly given that it is not quantified in any meaningful way. For instance, the authors could have run another correlation specifically on the data from May. Finally, the lack of maxima being evident at sunrise in their Figure 4 could be for multiple reasons, most notably the fact that sunrise times vary throughout the year [41]. In addition, sunset times also differ, along with daily maximum temperatures, and the combination of these factors as well as myriad others [42] affects timing and magnitude of night time temperature minima.

We are also somewhat concerned by the analyses that Ramsey et al. [1] ran on temperature and whooping signal occurrence. They chose to analyze their time-series using simple bivariate correlations, although cross-correlation would be more appropriate because this accounts for both non-independence and time lag [43].

### *Defining honey bee vibrational signals*

Honey bee communication, like nearly all animal communication [44], is highly variable and there is often no immediate and direct correspondence between a signal and a response. The stereotypy and information content of a signal like a waggle dance can also be quite variable, but there is, nonetheless, information communicated [45–47]. Even a sensory neuron will not always fire when appropriately stimulated [48]. Thus, to find a case in which one or a few examples do not meet expectations is not sufficient. Manipulative experiments need to be conducted and the results need to be quantitatively analysed, not qualitatively assessed.

However, we believe that the authors have contributed to our understanding of honey bee vibrational signalling. Automated sensing is an important innovation for the study of honey bee communication, though it is not a substitute for detailed behavioral observations that provide context and quantitative experiments that test hypotheses. We are uncertain if the authors have recorded stop signals, but we generally agree with their statement that stop signals are not exclusively waggle dance inhibitors (p 13, [1]). This was previously suggested by Thom et al. [19] for *A. mellifera* and demonstrated by Tan et al. [49] for *A. cerana* in context of nest attacks by a predator. Honey bee communication is complex, and many aspects, such as the relationship between tremble dancing and stop signalling [15,19,20], have not yet been elucidated. Researchers should ask how bees distinguish these different piping signals [reviewed in 50] and if they are acoustically different. We strongly suspect that context is important, as it is in most animal signals [44], and that these vibrational signals, like other honey bee signals, are rather variable.

*Acknowledgements*

The authors would like to thank Thomas Seeley for his thoughtful comments on our manuscript.